\documentstyle[12pt]{article}

\catcode`\@=11
\long\def\@makefntext#1{ 
\protect\noindent \hbox to 3.2pt {\hskip-.9pt
$^{{\ninerm\@thefnmark}}$\hfil}#1\hfill} 

\def\thefootnote{\fnsymbol{footnote}}
 \def\@makefnmark{\hbox to 0pt{$^{\@thefnmark}$\hss}}  

\def\ps@myheadings{\let\@mkboth\@gobbletwo
\def\@oddhead{\hbox{} 
\rightmark\hfil\ninerm\thepage}
\def\@oddfoot{}\def\@evenhead{\ninerm\thepage\hfil 
\leftmark\hbox{}}\def\@evenfoot{}
\def\sectionmark##1{}\def\subsectionmark##1{}}

\begin{document}

\newcommand{\half} {\mbox{\small $\frac{1}{2}$}}          

\newcommand{\symbolfootnote}{\renewcommand{\thefootnote}
        {\fnsymbol{footnote}}}
\renewcommand{\thefootnote}{\fnsymbol{footnote}}
\newcommand{\alphfootnote}
        {\setcounter{footnote}{0}
         \renewcommand{\thefootnote}{\sevenrm\alph{footnote}}}

\newcounter{sectionc}\newcounter{subsectionc}\newcounter{subsubsectionc}
\renewcommand{\section}[1] {\vspace{0.6cm}\addtocounter{sectionc}{1}
\setcounter{subsectionc}{0}\setcounter{subsubsectionc}{0}\noindent
        {\bf\thesectionc. #1}\par\vspace{0.4cm}}
\renewcommand{\subsection}[1] {\vspace{0.6cm}\addtocounter{subsectionc}{1}
        \setcounter{subsubsectionc}{0}\noindent
        {\it\thesectionc.\thesubsectionc. #1}\par\vspace{0.4cm}}
\renewcommand{\subsubsection}[1]
{\vspace{0.6cm}\addtocounter{subsubsectionc}{1}
        \noindent {\rm\thesectionc.\thesubsectionc.\thesubsubsectionc.
        #1}\par\vspace{0.4cm}}
\newcommand{\nonumsection}[1] {\vspace{0.6cm}\noindent{\bf #1}
        \par\vspace{0.4cm}}

\newcounter{appendixc}
\newcounter{subappendixc}[appendixc]
\newcounter{subsubappendixc}[subappendixc]
\renewcommand{\thesubappendixc}{\Alph{appendixc}.\arabic{subappendixc}}
\renewcommand{\thesubsubappendixc}
        {\Alph{appendixc}.\arabic{subappendixc}.\arabic{subsubappendixc}}

\renewcommand{\appendix}[1] {\vspace{0.6cm}
        \refstepcounter{appendixc}
        \setcounter{figure}{0}
        \setcounter{table}{0}
        \setcounter{equation}{0}
        \renewcommand{\thefigure}{\Alph{appendixc}.\arabic{figure}}
        \renewcommand{\thetable}{\Alph{appendixc}.\arabic{table}}
        \renewcommand{\theappendixc}{\Alph{appendixc}}
        \renewcommand{\theequation}{\Alph{appendixc}.\arabic{equation}}
        \noindent{\bf Appendix \theappendixc #1}\par\vspace{0.4cm}}
\newcommand{\subappendix}[1] {\vspace{0.6cm}
        \refstepcounter{subappendixc}
        \noindent{\bf Appendix \thesubappendixc. #1}\par\vspace{0.4cm}}
\newcommand{\subsubappendix}[1] {\vspace{0.6cm}
        \refstepcounter{subsubappendixc}
        \noindent{\it Appendix \thesubsubappendixc. #1}
        \par\vspace{0.4cm}}

\def\abstracts#1{{
        \centering{\begin{minipage}{30pc}\tenrm\baselineskip=12pt\noindent
        \centerline{\tenrm ABSTRACT}\vspace{0.3cm}
        \parindent=0pt #1
        \end{minipage} }\par}}

\newcommand{\bibit}{\it}
\newcommand{\bibbf}{\bf}
\renewenvironment{thebibliography}[1]
        {\begin{list}{\arabic{enumi}.}
        {\usecounter{enumi}\setlength{\parsep}{0pt}
\setlength{\leftmargin 1.25cm}{\rightmargin 0pt}
         \setlength{\itemsep}{0pt} \settowidth
        {\labelwidth}{#1.}\sloppy}}{\end{list}}

\topsep=0in\parsep=0in\itemsep=0in
\parindent=1.5pc

\newcounter{itemlistc}
\newcounter{romanlistc}
\newcounter{alphlistc}
\newcounter{arabiclistc}
\newenvironment{itemlist}
        {\setcounter{itemlistc}{0}
         \begin{list}{$\bullet$}
        {\usecounter{itemlistc}
         \setlength{\parsep}{0pt}
         \setlength{\itemsep}{0pt}}}{\end{list}}

\newenvironment{romanlist}
        {\setcounter{romanlistc}{0}
         \begin{list}{$($\roman{romanlistc}$)$}
        {\usecounter{romanlistc}
         \setlength{\parsep}{0pt}
         \setlength{\itemsep}{0pt}}}{\end{list}}

\newenvironment{alphlist}
        {\setcounter{alphlistc}{0}
         \begin{list}{$($\alph{alphlistc}$)$}
        {\usecounter{alphlistc}
         \setlength{\parsep}{0pt}
         \setlength{\itemsep}{0pt}}}{\end{list}}

\newenvironment{arabiclist}
        {\setcounter{arabiclistc}{0}
         \begin{list}{\arabic{arabiclistc}}
        {\usecounter{arabiclistc}
         \setlength{\parsep}{0pt}
         \setlength{\itemsep}{0pt}}}{\end{list}}

\newcommand{\fcaption}[1]{
        \refstepcounter{figure}
        \setbox\@tempboxa = \hbox{\tenrm Fig.~\thefigure. #1}
        \ifdim \wd\@tempboxa > 6in
           {\begin{center}
        \parbox{6in}{\tenrm\baselineskip=12pt Fig.~\thefigure. #1 }
            \end{center}}
        \else
             {\begin{center}
             {\tenrm Fig.~\thefigure. #1}
              \end{center}}
        \fi}

\newcommand{\tcaption}[1]{
        \refstepcounter{table}
        \setbox\@tempboxa = \hbox{\tenrm Table~\thetable. #1}
        \ifdim \wd\@tempboxa > 6in
           {\begin{center}
        \parbox{6in}{\tenrm\baselineskip=12pt Table~\thetable. #1 }
            \end{center}}
        \else
             {\begin{center}
             {\tenrm Table~\thetable. #1}
              \end{center}}
        \fi}

\def\@citex[#1]#2{\if@filesw\immediate\write\@auxout
        {\string\citation{#2}}\fi
\def\@citea{}\@cite{\@for\@citeb:=#2\do
        {\@citea\def\@citea{,}\@ifundefined
        {b@\@citeb}{{\bf ?}\@warning
        {Citation `\@citeb' on page \thepage \space undefined}}
        {\csname b@\@citeb\endcsname}}}{#1}}

\newif\if@cghi
\def\cite{\@cghitrue\@ifnextchar [{\@tempswatrue
        \@citex}{\@tempswafalse\@citex[]}}
\def\citelow{\@cghifalse\@ifnextchar [{\@tempswatrue
        \@citex}{\@tempswafalse\@citex[]}}
\def\@cite#1#2{{$\null^{#1}$\if@tempswa\typeout
        {IJCGA warning: optional citation argument
        ignored: `#2'} \fi}}
\newcommand{\citeup}{\cite}

\def\fnm#1{$^{\mbox{\scriptsize #1}}$}
\def\fnt#1#2{\footnotetext{\kern-.3em
        {$^{\mbox{\sevenrm #1}}$}{#2}}}

\font\twelvebf=cmbx10 scaled\magstep 1
\font\twelverm=cmr10 scaled\magstep 1
\font\twelveit=cmti10 scaled\magstep 1
\font\elevenbfit=cmbxti10 scaled\magstephalf
\font\elevenbf=cmbx10 scaled\magstephalf
\font\elevenrm=cmr10 scaled\magstephalf
\font\elevenit=cmti10 scaled\magstephalf
\font\bfit=cmbxti10
\font\tenbf=cmbx10
\font\tenrm=cmr10
\font\tenit=cmti10
\font\ninebf=cmbx9
\font\ninerm=cmr9
\font\nineit=cmti9
\font\eightbf=cmbx8
\font\eightrm=cmr8
\font\eightit=cmti8


\centerline{\tenbf THE ``SPIN'' STRUCTURE OF THE NUCLEON%
            \footnote{Talk given by R. Horsley at the $2^{nd}$
                      IMACS Conference on Computational Physics,
                      Cahokia, USA.}}
\baselineskip=16pt
\centerline{\tenbf -- A LATTICE INVESTIGATION}
\vspace{0.8cm}
\centerline{\tenrm R. ALTMEYER, G. SCHIERHOLZ}
\baselineskip=13pt
\centerline{\tenit DESY, Notkestra{\ss}e 85, D-22603 Hamburg, Germany}
\vspace{0.3cm}
\centerline{\tenrm M. G\"OCKELER, R. HORSLEY}
\baselineskip=13pt
\centerline{\tenit HLRZ, c/o Forschungszentrum J{\"u}lich,
                   D-52425 J{\"u}lich, Germany}
\vspace{0.3cm}
\centerline{\tenrm and}
\vspace{0.3cm}
\centerline{\tenrm E. LAERMANN}
\baselineskip=13pt
\centerline{\tenit Fakult{\"a}t f{\"u}r Physik, Universit{\"a}t Bielefeld,
                   D-33501 Bielefeld, Germany}
\vspace{0.9cm}
\abstracts{We will discuss here an indirect lattice evaluation of the
baryon axial singlet current matrix element. This quantity may be related to
the fraction of nucleon spin carried by the quarks. The appropriate
structure function has recently been measured (EMC experiment).
As in this experiment, we find that the quarks do not appear to carry
a large component of the nucleon spin.}

\twelverm   
\baselineskip=14pt
\section{Introduction and Theoretical Discussion}
\label{theory}
Hadrons appear to be far more complicated than the (rather successful)
constituent quark model would suggest. For example an old result is
from the $\pi N$ sigma term, which seems to give a
rather large strange component to the nucleon mass. A more recent
result is from the EMC experiment\cite{ashman89a} which suggests that
constituent quarks are responsible for very little of the nucleon spin.
These are non-perturbative effects and so is an area where lattice
calculations may be of some help.

The EMC experiment measured deep inelastic scattering (DIS) using a
$\mu$ beam on a proton target. The new element in the experiment was
that both the inital $\mu$ and proton were longitudinally polarised.
A measurement of the difference in the cross sections for parallel and
anti-parallel polarised protons enabled the structure function
$g_1(x,Q^2)$ to be found. Theoretically\cite{bass92a} this is of
interest as using the Wilson operator product expansion,
moments of the structure function are related to certain matrix elements.
In this case, defining
$s^\mu \Delta q = \langle Ps|\bar{q}\gamma^\mu\gamma_5 q|Ps\rangle$
for $q = u$, $d$ or $s$ quarks (i.e. the expectation value of the axial
singlet current) the lowest moment is then given by
\noindent \raisebox{19.5cm}[0.0cm][0.0cm]{\hspace{-3.5cm}
                                          Preprint HLRZ 93-71 \hspace{0.25cm}
                                          November 1993}
\begin{eqnarray}
   \int_0^1 \! dx g_1(x,Q^2)
         &=& {1\over 2} \left( {4\over9} \Delta u + {1\over9} \Delta d +
                               {1\over9} \Delta s \right) \nonumber \\
         &\approx&  0.20 + {1\over 3}\Delta s ,
\label{theory.a}
\end{eqnarray}
($Q^2_{EMC} \approx 11 \mbox{GeV}^2$). Results from neutron and hyperon
decays have been used to eliminate two of the unknowns on the RHS of
this equation.

$\Delta q$ can be given a physical interpretation, as the quark spin
operator, $\hat{S}_i^{quark}$, is local and gauge invariant
(in distinction to the gluon spin operator or orbital angular momentum
operator)\cite{jaffe90a} and leads to
\begin{eqnarray}
   2 \langle Ps| \hat{S}_i^{quark} |Ps \rangle \equiv \Delta \Sigma
       &=& \Delta u + \Delta d + \Delta s   \nonumber \\
       &\approx& 0.68 + 3\Delta s.
\label{theory.b}
\end{eqnarray}
Before the EMC experiment, the Ellis-Jaffe sum rule was applied:
the strange content of the proton was taken as zero, i.e. $\Delta s =0$.
However the EMC experiment gave for the LHS of eq.~(\ref{theory.a})
$\approx 0.126$ giving $\Delta s \approx -0.2$, a large negative strange
contribution to the proton and $\Sigma \approx 0$, i.e. the total
quark spin is zero. This situation is often referred to as
``the spin crisis of the quark model'', although what was actually
measured was the nucleon matrix element of the axial singlet current.

\section{The lattice calculation}
\label{lattice}

We now turn to our lattice calculation. By working in euclidean space
on a lattice we can turn our problem into a statistical mechanics
one, which can be approached numerically via the calculation of
correlation functions using Monte Carlo techniques. We have generated
configurations\cite{altmeyer93a} using dynamical staggered fermions
($\chi$) on a $16^3 \times 24$ lattice at $\beta =5.35$,
$m=0.01$, which corresponds to a quark mass of about $35 \mbox{MeV}$.
Staggered fermions describe $4$ degenerate quark flavours,
so we do not have quite the physical situation that we wish to describe
(i.e. $2$ light quarks and one heavier quark).
Nevertheless this discretisation of the fermions has some
advantages -- most notably good chiral properties (as $m \to 0$)
when the chiral symmetry is spontaneously broken and there is a
Goldstone boson, the $\pi$.

The general procedure on the lattice to measure matrix elements
is to consider $3$-point correlation functions\cite{maiani87a}.
For an arbitrary operator $\Omega$, it can then be shown that
($B$ is the nucleon operator)
\begin{eqnarray}
   C(t;\tau) &=& \langle B_{\vec{p},\vec{s}}(t) \Omega(\tau)
                 \bar{B}^\prime_{\vec{p}^{\,\prime}, \vec{s}^{\,\prime}}(0)
                 \rangle                    \nonumber \\
             &=& \sum_{\alpha,\beta =N, \Lambda} \! A_{\alpha\beta}
                 \langle \alpha | \hat{\Omega} | \beta \rangle
                 \mu_\alpha^{t-\tau} \mu_\beta^\tau ,
\label{lattice.a}
\end{eqnarray}
for $\half T \gg t \gg \tau \gg 0$, where $T$ is the time box size
$= 24$ here. $\mu_N = + \exp{(-E_N)}$ and the parity partner
(which always occurs when using staggered fermions) is denoted
by $\Lambda$, so that $\mu_\Lambda = - \exp{(-E_\Lambda)}$.
The amplitudes $A_{\alpha\beta}$ and energies $E_\alpha$ are known
from $2$-point correlation functions. Thus from eq.~(\ref{lattice.a})
we can extract $\langle Ps|\hat{\Omega}|Ps\rangle$. Simply setting $\Omega$
to be $\bar{\chi} \gamma_i\gamma_5 \chi$ has, however, a number of
disadvantages: the correlation function in eq.~(\ref{lattice.a})
has connected and disconnected parts, in the fit there are cross
terms present, and the operator must be renormalised. Another
approach is to consider the divergence of the current\cite{mandula92a}.
This can be related to the QCD anomaly. In the chiral limit an equivalent
formulation of the problem is thus ($n_f =4$)
\begin{equation}
   \Delta\Sigma = \lim_{\vec{p}^{\,\prime} \to \vec{p}}
                  {n_f\over {i(\vec{p}-\vec{p}^{\,\prime})\cdot \vec{s}}}
                  \langle P(\vec{p}),\vec{s}|
                  \hat{Q}(\vec{p}-\vec{p}^{\,\prime})
                  | P(\vec{p}^{\,\prime}), \vec{s}^{\,\prime} \rangle ,
\label{lattice.b}
\end{equation}
with topological charge
\begin{equation}
   \hat{Q}(\vec{p}) = {1 \over {8\pi^2}}
                       \int \! d^3\vec{x} F_{\mu\nu}\tilde{F}^{\mu\nu}
                       e^{i\vec{p}\cdot \vec{x}} .
\label{lattice.c}
\end{equation}
(We shall take $\vec{p}^{\,\prime} = \vec{0}$, $\vec{p} =(0,0,p)$.)
On the lattice we have used the L\"uscher\cite{luescher82a} construction for
this charge. Although technically complicated it is integer valued
and hence has no renormalisation. In the $3$-point correlation function
there are no cross terms ($Q$ has a definite parity). A disadvantage is that
we need to take $p \to 0$. On our lattice the smallest momentum available
is the rather large $p \approx 500\mbox{MeV}$.

Nevertheless on attempting this measurement, we find a reasonable
signal with $\Delta\Sigma \approx 0.18(2)$. This is
small and tends to support the EMC result, which would indicate
a rather large sea contribution to the proton. The existence of the
QCD anomaly also proved important. Further details of our calculation
are given in\cite{altmeyer92a}.

\section{Acknowledgements}
\label{acknowledgements}
This work was supported in part by the DFG. The numerical simulations
were performed on the HLRZ Cray Y-MP in J\"ulich. We wish to thank
both institutions for their support. The configurations used were those
generated for the $MT_c$ project.

\section{References}
\label{references}

\end{document}